\documentstyle[sprocl]{article}

\input{psfig}
\def\epem{$e^+e^-$}
\def\ee{e^+e^-}
\def\chp#1{\widetilde\chi^{~+}_{#1}}
\def\chm#1{\widetilde\chi^{~-}_{#1}}
\def\chpm#1{\widetilde\chi^{~\pm}_{#1}}
\def\chmp#1{\widetilde\chi^{~\mp}_{#1}}
\def\chz#1{\widetilde\chi^{~0}_{#1}}

\def\sell{\tilde e_L}
\def\selr{\tilde e_R}

\def\smupl{\widetilde \mu^{~+}_L}
\def\smuml{\widetilde \mu^{~-}_L}
\def\smupr{\widetilde \mu^{~+}_R}
\def\smumr{\widetilde \mu^{~-}_R}
\def\smul{\widetilde \mu_L}
\def\smur{\widetilde \mu_R}

\def\stau#1{\widetilde \tau_{#1}}

\def\snu{\widetilde \nu}
\def\snue{\widetilde \nu_e}
\def\snum{\widetilde \nu_\mu}
\def\snut{\widetilde \nu_\tau}

\def\ifb{fb$^{-1}$}

\def\me{\mbox{$\rlap{\kern0.2em/}E$}}

\begin{document}

\title{A RUN SCENARIO FOR THE LINEAR COLLIDER}

\author{ PAUL D. GRANNIS }

\address{Department of Physics and Astronomy, State University of New York,\\
Stony Brook NY 11790, USA}

\maketitle\abstracts{
We outline a run plan and 
examine the precision with which supersymmetric and Higgs parameters
may be determined in 1000 fb$^{-1}$ of running at the nominal 500 GeV 
\epem ~linear collider, assuming the Snowmass benchmark mSUGRA point SM2
(similar to the SPS1 benchmark).
}
  
\section{Introduction}

The physics program of the linear \epem ~collider (LC) can 
be very rich, particularly in the case that low mass supersymmetry exists.
We examine a run plan for such a physics-rich case,
in which the energies of the LC are selected to
first measure the sparticle masses approximately using energy distributions
at the highest available energy, followed by a series of 
scans at selected superparticle pair thresholds.  In the supersymmetry
part of this study, we pay attention to the way in which one can
identify specific sparticle states based on the 
distinctive final states with little background. This paper is an update and
extension of the study reported in the 2001 Snowmass Workshop proceedings
\cite{snowmass}.

We assume a physics scenario in which supersymmetry is
characterized by the minimal supergravity model corresponding
to the Snowmass SM2 benchmark ($m_0=100$ GeV, $m_{1/2}=250$ GeV, 
$\tan\beta=10$, $A_0=0$, and sign$\mu=+$).  This benchmark point
has nearly the same sparticle mass and branching ratios as the 
subsequent SPS1a point~\cite{allanach}.  We take the lightest Higgs boson mass
to be 120 GeV, and to have standard model (SM) properties.

We have considered the initial
1000 fb$^{-1}$ of LC running at 500 GeV, expected to take roughly the first
seven years of operation. For runs below 500 GeV, we assume the
luminosity falls as $1/E$. 
We assume that the LC operates with either left- or right-handed electron
polarization of 80\% and assume no positron polarization.
 The proposed runs and associated polarization
states are given in 
Table~\ref{tab:runsps1}. 
When a run with left or right polarization is called for, we assume
that about 90\% of the data is taken with the specified polarization
and about 10\% with the other polarization.
  This run plan differs from that in
Ref.[1] in that there is no dominantly right polarization run
at 410 GeV, and there is added running at the $Z$ boson mass.

\begin{table}[t]
\caption{Run allocations for the SPS1 Minimal Sugra parameters. 
\label{tab:runsps1}}
\vspace{0.2cm}
\begin{center}
\begin{tabular}{|cccccl|}
\hline
Beams & Energy & Pol. & $\int {\cal L}dt$ 
  & [$\int {\cal L}dt]_{\rm equiv}$ & ~~~~~~~~Comments \\  \hline
$\ee $ & 500 & L/R & 335 & 335 & Sit at top energy for sparticle
       masses \\ \hline
$\ee $ & $M_Z$ & L/R & 10 & 45 & Calibrate with $Z$'s \\ \hline
$\ee $ & 270 & L/R & 100 & 185 & Scan $\chz1~\chz2$ 
                                        threshold (L pol.) \\
       &     &     &     &     & Scan $\stau1~\stau1$ threshold 
                                        (R pol.) \\ \hline
$\ee $ & 285 & R & 50 & 85 & Scan $\smupr~\smumr$ threshold \\ \hline
$\ee $ & 350 & L/R & 40 & 60 & Scan $t\overline t$ threshold\\ 
       &     &     &     &     & Scan $\selr~\sell$ threshold 
                                      (L \& R pol.) \\
       &     &     &     &     & Scan $\chp1~\chm1$ threshold (L pol.) \\
                                        \hline
$\ee $ & 410 & L & 60 & 75 & Scan $\stau2~\stau2$ threshold \\ 
      &      &   &    &    & Scan $\smupl~\smuml$ threshold \\ \hline
$\ee $ & 580 & L/R & 90 & 120 & Sit above $\chpm1~\chmp2$ threshold for 
	                           $\chpm2$ mass\\ \hline \hline
$e^-e^- $ & 285 & RR & 10 & 95 & Scan with $e^-e^-$ collisions for 
                   $\selr$ mass   \\ \hline
\end{tabular}
\end{center}
\end{table}

\section{Initial estimates of sparticle masses from end points}

For the SM2 Susy benchmark, the sparticle masses and primary 
decay branching ratios are as given in Table~\ref{tab:sps1mass}.
The $h^0$, $H^0$, $A^0$ and $H^\pm$ masses are 113, 380,
379 and 388 GeV respectively.  The spartners of the lighter
quarks and the gluino have masses of about 530 and 595 GeV.
The two top squark masses are 393 and 572 GeV.

Although many particles are accessible at 500 GeV,
the states $\stau1$, $\stau2$, $\chz2$, $\chpm1$ have
dominant decays into $\tau$ final states. In addition, the
sneutrinos have dominant wholly invisible decays.
On the other hand, the $\chz3$ has distinctive decays
involving $Z$ bosons.   Such observations are indicative of the fact that
the Susy spectra are likely to vary drastically
as parameters are changed, and thus specific
conclusions on the precisions obtainable for sparticle properties
in one scenario should not be blindly translated to others.

\begin{table}[t]
\caption{Masses (in GeV) and dominant branching fractions for the SM2
benchmark.\label{tab:sps1mass}}
\vspace{0.2cm}
\begin{center}
\begin{tabular}{|c|c|lllll|}
\hline
~ & M & ~~Final state &(BR(\%)) & ~ & ~ & \\ \hline
$\selr$ & 143 & $\chz1 e$ (100)~& & & &  \\
$\sell$ & 202 & $\chz1 e$ (45)~ & 
                  $\chpm1 \nu_e$ (34)~ &
                  $\chz2 e$ (20)~ & & \\
$\smur$ & 143 & $\chz1 \mu$ (100)~ & & & & \\
$\smul$ & 202 & $\chz1 \mu$ (45)~ & 
                  $\chpm1 \nu_\mu$ (34)~ &
                  $\chz2 \mu$ (20)~ & & \\
$\stau1$ & 135 & $\chz1 \tau$ (100)~ & & & & \\
$\stau2$ & 206 & $\chz1 \tau$ (49)~ & 
                  $\chm1 \nu_\tau$ (32)~ &
                  $\chz2 \tau$ (19)~ & & \\ \hline
$\snue$ & 186 & $\chz1 \nu_e$ (85)~ & 
                  $\chpm1 e^\mp$ (11)~ &
                  $\chz2 \nu_e$ (4)~ & & \\
$\snum$ & 186 & $\chz1 \nu_\mu$ (85)~ & 
                  $\chpm1 \mu^\mp$ (11)~ &
                  $\chz2 \nu_\mu$ (4)~ & & \\ 
$\snut$ & 185 & $\chz1 \nu_\tau$ (86)~ & 
                  $\chpm1 \tau^\mp$ (10)~ &
                  $\chz2 \nu_\tau$ (4)~ & & \\ \hline
$\chz1$ & 96 & stable~ & & & & \\
$\chz2$ & 175 & $\stau1 \tau$ (83)~ &
                  $\selr e$ (8)~ & 
                  $\smur \mu$ (8)~ & & \\
$\chz3$ & 343 & $\chpm1 W^\mp$ (59)~ & 
                  $\chz2 Z$ (21)~ & 
                  $\chz1 Z$ (12)~ & 
                  $\chz1 h$ (2)~ &  \\
$\chz4$ & 364 & $\chpm1 W^\mp$ (52)~ & 
                  $\snu \nu$ (17)~  &
                  $\stau2 \tau$ (3)~ & 
                  $\widetilde \chi_{1,2} Z$ (4)~ &
                  $\widetilde\ell_R \ell$ (6)~ \\ \hline
$\chpm1$ & 175 & $\stau1 \tau$ (97)~ & 
                  $\chz1 q\overline q$ (2)~ & 
                  $\chz1 \ell\nu$ (1.2)~ & & ~ \\
$\chpm2$ & 364 & $\chz2 W$ (29)~ &
                 $\chpm1 Z$ (24)~ &
                   $\widetilde\ell \nu_\ell$ (18)~ &
                   $\chpm1 h$ (15)~ &
                   $\snu_\ell \ell$ (8)~ \\ \hline
\end{tabular}
\end{center}
\end{table}

The traditional `end point' mass determinations~\cite{martyn,nauenberg}
seek the sharp edges
in the energy distribution of a SM particle $C$ in the 
two body sparticle decay
$\widetilde A \rightarrow \widetilde B C$.  
Such sharp edges require
that a given final state channel under observation be fed mainly by
just one Susy reaction, and that the SM particle $C$ be stable.  Neither of
these conditions hold for many of the channels involved in our benchmark Susy
point.  The second condition is of course not actually required, since any {\it
known} kinematic distribution can be fit with a set of templates derived for
a set of hypothesized masses, and the best mass obtained from the fits.  The
problem of multiple reactions feeding a particular observed channel is more
serious, and must be addressed in detail for any given Susy benchmark.  We
comment here on the likely ability to determine sparticle masses for the SM2
benchmark.  Our estimates of mass precision have been guided by previous
studies~\cite{martyn,nauenberg}, 
and we scale errors according to the ratio of our event samples relative
to these studies.

We restrict attention to channels in which there are only leptons
and missing energy, for which standard model backgrounds should be
of little importance.  
We have followed the full decay chains of those sparticle pair
reactions accessible at 500 GeV, weighted by their production cross sections
and decay branching ratios.  The number of events contributed 
to all 2, 4 and 6 lepton final states from all reactions that feed the channel
were tabulated for each initial electron polarization state.  
No SM backgrounds are assumed.  We have looked for the channel/polarization
combination in which the production of a specific Susy particle are most
dominant; in some cases there is no wholly dominant channel, and we imagine
that a set of coupled-channel studies will be needed to get several sparticle
masses simultaneously.  We discuss the specific 
sparticle template mass determinations in turn.

The $\smur$ and $\smul$ masses are simple to obtain.  The final state
$\mu^+\mu^-$\me ~with right $e^-$ polarization has 95\% of its contribution
from 
$\smur^\pm \smur^\mp$ production, and the decay $\smur \rightarrow
\chz1 \mu$ decay determines both the $\chz1$ and $\smur$ masses.   
The final state $\mu^\pm \tau^\mp$\me
~final state from left polarized electrons is 95\% due to $\smul^\pm 
\smul^\mp$ production, with one $\smul$ decay to $\chz1 \mu$
and the other $\smul$ to 
$\nu_\mu \chpm1 \rightarrow \nu_\mu \nu_\tau \tau \chz1$.
The $\mu$ energy distribution gives $M(\smul)$ in this case.

The $e^+e^-$\me ~final state is dominated by the production of
$\selr^+ \selr^-$, $\selr^+ \sell^-$, $\sell^+ \selr^-$ and $\sell^+ \sell^-$.
A Snowmass study~\cite{nauenberg} 
has shown that through the double difference of the
distributions for $e^+$ and $e^-$ end points with 
$L$- and $R$-polarized electrons, the masses of $\selr$, $\sell$
(and $\chz1$) can be well determined.

\begin{figure}
\psfig{figure=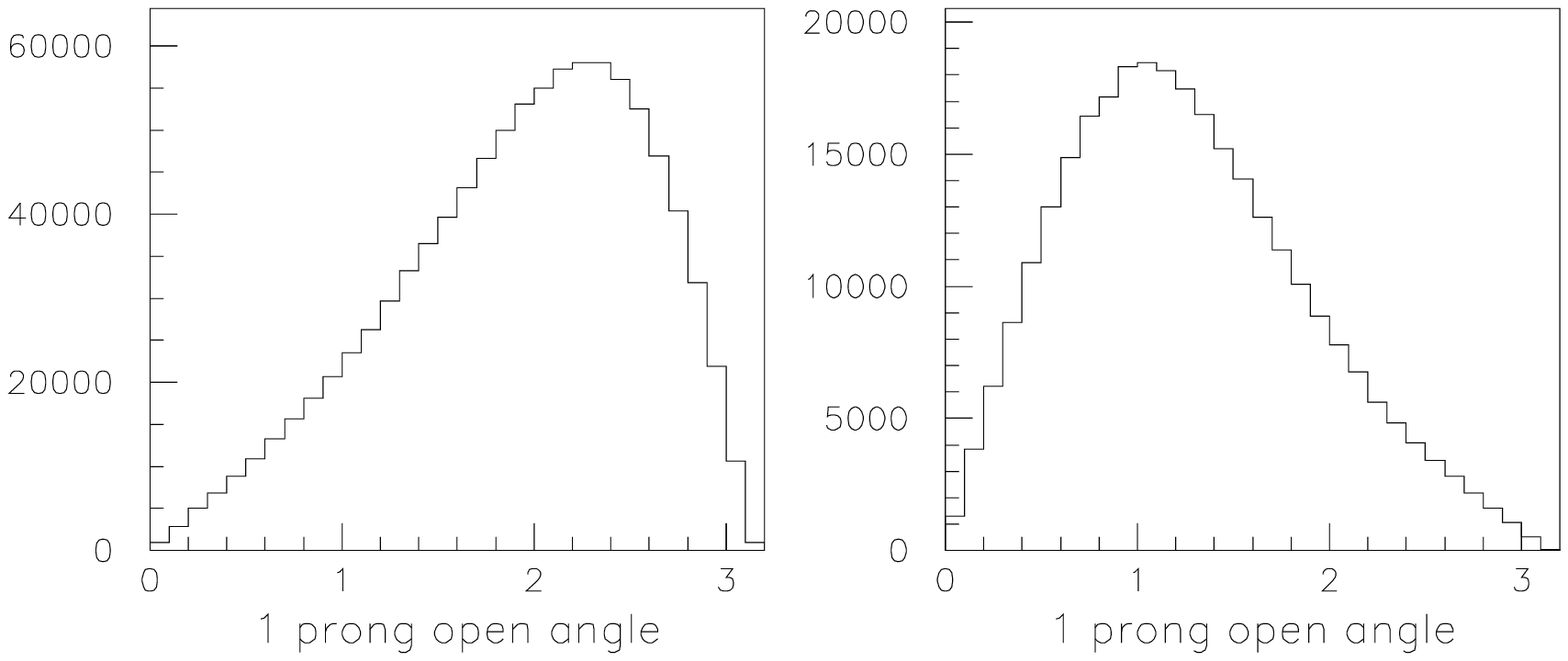,height=2.0in}
\caption{Opening angle of 1-prongs from $\tau$ decay (a) from reaction
$\chp1 \chm1$ and (b) $\chz1 \chz2$ from left-polarized electrons.
\label{fig:openang}}
\end{figure}

In benchmark SM2, the $\tau^+\tau^-$ final states comingle many
production processes.  Table~\ref{tab:tautau} shows the
number of events from various 
contributing reactions for both electron polarizations.
The $\chpm1, \chz2$, and $\stau1$ are all large contributors,
so their masses must be extracted together.  There are however
some added features that help disentangle the reactions. As shown
in Fig.~\ref{fig:openang}, for
the $\chp1 \chm1$ and $\stau1^+ \stau1^-$, the two $\tau$'s
tend to be back to back, whereas for $\chz1 \chz2$, the two $\tau$'s
come from the same parent $\chz2$ and tend to be more collinear.

\begin{table}[t]
\caption{Reactions feeding the $\tau \tau$\me ~final state
and the expected number of events before acceptance and efficiency
losses in 335 fb$^{-1}$.
\label{tab:tautau}}
\vspace{0.2cm}
\begin{center}
\begin{tabular}{|l|r|r|}
\hline
Reaction & N($e_L$)  & N($e_R$) \\  \hline
$\chp1 \chm1$ & 97,440 & 11,229 \\
$\chz1 \chz2$ & 29,424 & 6,846 \\
$\stau1^+ \stau1^-$ & 11,792 & 29,547 \\
$\stau2^+ \stau2^-$ (via $\chz1 \tau$) & 5,716 & 2,027 \\
$\sell^+ \sell^-$ & 3,905 & 625 \\
$\smul^+ \smul^-$  & 1,395 & 428  \\
$\stau2^+ \stau2^-$ (via $\chpm1 \nu_\tau$)  & 1,004 & 356  \\
$\stau1^\pm \stau2^\mp$  & 805 & 644  \\
$\chz1 \chz3$  & 71 & 85  \\  \hline
\end{tabular}
\end{center}
\end{table}

For the three dominant reactions feeding the 
left-polarized $\tau \tau$\me ~channel,
the sparticles involved in specifying the energy distributions are
the well-measured $\chz1, \widetilde e,$ and $\widetilde \mu$, together with
the $\chpm1, \chz2$ and $\stau1$ whose masses we want to estimate.
The energy distributions of the one-prongs from $\tau$ decays 
are not box-like, but are as shown in 
Fig.~\ref{fig:eprong}.  These distributions can be fit however for
the unknown sparticle masses using templates.  
For example, with the available statistics in
335 \ifb, we find that the statistical error on the $\stau1$ mass is 0.22 GeV
when fitting such templates
(keeping the $\chpm1$ and $\chz2$ masses fixed).
In a real analysis, all three masses must of course be varied, but more channels
may be employed.

\begin{figure}
\psfig{figure=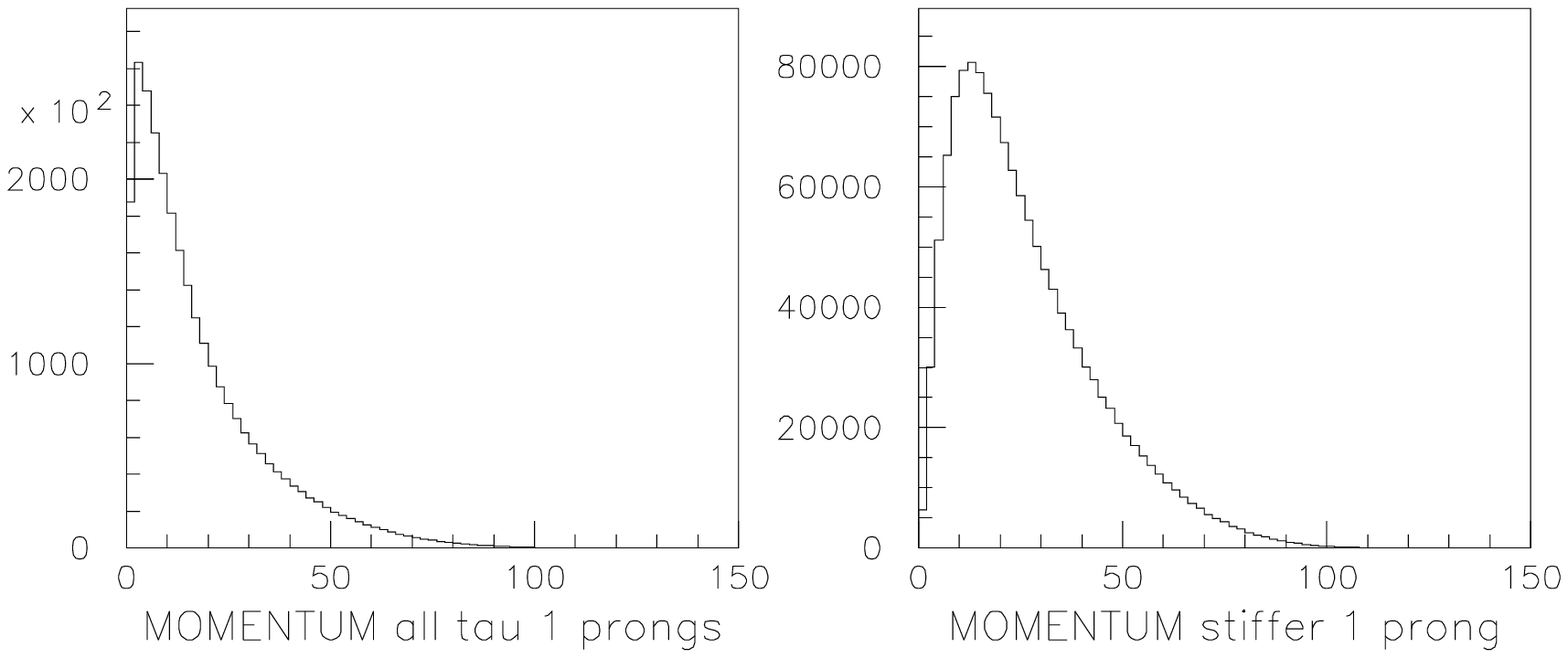,height=2.0in}
\caption{Energy distributions of 1-prongs from $\tau$ decay in 
the $\tau^+\tau^-$ channel; (a) all 1-prongs, and
(b) the more energetic 1-prong.
\label{fig:eprong}}
\end{figure}

Despite the fact that no single reaction dominates a particular channel, there
are several channels that are dominated by the same unknown
$\chpm1, \chz2$ and $\stau1$ masses in different proportions.
For example the $\mu$ in the $\mu\tau\tau\tau$ channel with left polarization
arises
from $\smul \rightarrow \mu\chz2$ 92\% of the time (there are 1400 events before
efficiency losses), so gives a reasonable determination of the $\chz2$ mass.
Other channels involving the same three sparticles are 
$\tau \tau$ (right polarized), $\mu\mu\tau\tau$ (left),
$\mu\tau\tau\tau$ (left) and $\tau\tau\tau\tau$ (left).
A careful study is needed to see how well the three unknown masses can
be disentangled from a coupled channel analysis.  I guess that
precisions of about 1 GeV should be achievable from the 500 GeV
runs, sufficient to locate the energy at which a scan should be performed.

The higher mass gaugino masses may be partly measurable.
The $\chz3$ has the useful and distinguishable 
decay modes, $\chz3 \rightarrow \chz1 (\chz2)
Z$ from which the $\chz3$ mass can be determined, albeit with limited
statistics.  The $\chz4$ mass though accessible kinematically, has 
event rates that are too small 
for a mass measurement.  The $\chpm2 \chmp1$ reaction threshold
opens at 539 GeV in the SM2 benchmark; if we can deduce that this is is
likely to be the case, we would raise the LC energy to about 580 GeV
to allow $\chpm2 \chmp1$ production and give a rough estimate of
the $\chpm2$ mass.

The $\widetilde \nu$ masses are difficult to obtain in this scenario.
The most favorable channel for isolating $\snue$ pair production 
is the $e\tau\tau\tau$ final state, where $\snue \snue^*$ contributes
about 40\% of the 6.5K events (before efficiency losses).
The $ee\tau\tau$ and $e\mu\mu\tau$ channels have significant
$\snue \snue^*$ contributions.  Perhaps it is possible to dig the
$\snue$ mass out.  The $\snum$ and $\snut$ seem hopeless to measure,
as these sparticles do not come close to dominating any final state
channel.

\section{Sparticle threshold scans}

Previous studies~\cite{martyn} 
have examined the precision that can be attained on
mass measurements from the appropriate energy scans near a two particle
threshold.  For this measurement, the problem of disentangling the
reaction feeding a particular final state does not occur, providing
that there are no overlapping thresholds in a particular channel.  However,
cross-sections near thresholds tend to be small, so the scans can use
rather significant portions of the luminosity budget, and should be chosen for
only those cases where there are substantial gains to be made.

The first studies~\cite{martyn} of scans used ten energy points
across the threshold in question.  Subsequent 
studies~\cite{mizukoshi,blair} have questioned this
strategy, finding that particularly in the case of low $\sigma\times$BR and
relatively slow threshold turn-ons characteristic of sfermion pairs 
($\beta^3$), fewer points on the scan may be better.  An analytic 
study~\cite{snowmass}
that did not include the effects of backgrounds showed that the 
strategy for minimizing the errors shows little improvement for
more than about three points on the scan.

One can improve the slope of the threshold curve for $\selr \selr$
by producing them in $s-$wave from an $e^-e^-$ initial state
(rather than in $p-$wave as for \epem) even
after the inclusion of beamsstrahlung effects and the reduced 
luminosity in $e^-e^-$ operation~\cite{feng}.  We have adopted
this strategy in our run plan.    

The estimated precisions
were scaled from those of previous threshold scan studies.  We have not 
explicitly included the effects arising from differences in background at our
benchmark point relative to previous studies.  However, we
conservatively considered only the dominant channel and polarization state.

\section{Susy mass and parameter precisions attainable}

The mass precisions obtained for the run plan of Table~\ref{tab:runsps1}
are shown in Table~\ref{tab:precision}.
Given these mass precisions and making the assumption that
the supersymmetry is indeed mSUGRA, the errors that can be obtained
for the Susy parameters are shown in Table ~\ref{tab:sugra}.

\begin{table}[t]
\caption{Mass precisions for the run plan of Table~\ref{tab:runsps1}.
\label{tab:precision}}
\vspace{0.2cm}
\begin{center}
\begin{tabular}{|c|cc|c|}
\hline
sparticle & $\delta M$ & $\delta M$  & $\delta M$  \\  
          &  end point & scan      & combined  \\ \hline
$\selr$ & 0.19 & 0.02  & 0.02 \\
$\sell$ & 0.27 & 0.30  & 0.20  \\
$\smur$ & 0.08 & 0.13  & 0.07  \\
$\smul$ & 0.70 & 0.76  & 0.51  \\
$\stau1$ & $\sim 1 - 2$ & 0.64  & 0.64  \\
$\stau2$ & -- & 1.1 & 1.1  \\
$\snue$ & $\sim 1$ & --  & $\sim 1$  \\
$\snum$ & 7? & --  & 7? \\
$\snut$ & -- & --  & --  \\
$\chz1$ & 0.07 & --  & 0.07  \\
$\chz2$ & $\sim 1 - 2$ & 0.12  & 0.12 \\
$\chz3$ & 8.5 & --  & 8.5  \\
$\chz4$ & -- & --  & --  \\
$\chpm1$ & $\sim 1-2$ & 0.18  & 0.18  \\
$\chpm2$ & 4 & --  & 4  \\ \hline
\end{tabular}
\end{center}
\end{table}

\begin{table}[t]
\caption{Errors on mSUGRA parameters.
\label{tab:sugra}}
\vspace{0.2cm}
\begin{center}
\begin{tabular}{|c|c|}
\hline
parameter & SPS1  \\ \hline
$m_0$ & $100 \pm 0.08$ GeV ~~~ \\
$m_{1/2}$ & $250 \pm 0.20$ GeV ~~~ \\
$A_0$ & $0 \pm 13$ GeV ~~~  \\
$\tan\beta$ & $10 \pm 0.47$ ~~~  \\ \hline
\end{tabular}
\end{center}
\end{table}

\section{Higgs and top quark studies}

The Higgs production processes occur at most of the energies
in our run plan; there as many $ZH$ events over the full run
as would be obtained in about 525 \ifb of running at 350 GeV,
or in about 1220 \ifb at 500 GeV.  We do not expect that running
at several energies should materially affect the Higgs parameter
determinations, and we simply scale the precision obtained in previous
studies~\cite{tesla,orange} by the statistics of the $ZH$ sample.
We have used the somewhat more favorable estimates for thge TESLA
study~\cite{tesla},
but note that there is still some uncertainty on the branching
ratio errors, in particular for $c\overline c$ decays.  The 
projected Higgs parameter errors are shown in Table~\ref{tab:higgs}.
Use of Higgs bosons produced through $WW$ fusion may improve the
situation somewhat.

The top quark parameters are taken from the scan near the $t\overline t$ 
threshold at 350 GeV.  The statistical errors are expected to be small compared
with the theoretical errors associated with QCD theory.
We may expect that the mass and width of the top quark may be determined
to 150 MeV and about 70 MeV respectively in the 40 \ifb allocated to
this threshold.  (Less luminosity 
would be needed to saturate expected theory errors for the top quark,
but the $t \overline t$ threshold overlaps those of $\sell \selr$ 
and $\chp1 \chm1$, for which larger luminosity accumulations are desired.)

\section{Calorimeter calibrations}

It will be necessary to make good 
relative calibrations of the calorimeter energy
scale in a LC detector, so as to capitalize upon the excellent intrinsic
resolution of the calorimeter.  These should be done with a
periodicity that is shorter than the gain-drift time for the calorimeter.
Two methods could be envisioned:  the first employing special runs
at the $Z$ boson mass, and the second using data collected during 
ordinary running at higher energies.

\begin{table}[t]
\caption{\label{tab:higgs}
Relative errors (in \%) on Higgs mass, cross-section,
total width, branching ratios and Yukawa couplings ($\lambda$)
for the run plan of Table \ref{tab:runsps1}.
}
\begin{center}
\begin{tabular}{|cc|cc|}
\hline 
Parameter & error & Parameter & error \\ \hline
Mass    & 0.03       & 
       $\Gamma_{\rm tot}$ & 7 \\
$\sigma$($ZH$) &  3  & 
       $\lambda_{ZZH}$   & 1     \\
$\sigma$($WW$) & 3   & 
       $\lambda_{WWH}$   & 1     \\
BR($b\overline b$) & 2 & 
       $\lambda_{bbH}$   & 2      \\
BR($c\overline c$) & 8 & 
       $\lambda_{ccH}$   & 4      \\
BR($\tau^+\tau^-$) & 5 & 
       $\lambda_{\tau\tau H}$ & 2      \\
BR($gg$)          & 5 & 
       $\lambda_{ttH}$   & 30    \\ \hline
\end{tabular}
\end{center}
\end{table}
 
Our run plan in Table~\ref{tab:runsps1} calls for four runs of 2 \ifb
each at the $Z$ (every other year).  
Others would ask for more -- perhaps
one per year.  For each such run, we expect 3.6 million $Z\rightarrow ee
~{\rm or}~\mu\mu$ and 76 million $Z\rightarrow$ hadrons.  We assume 
a calorimeter
with of order $10^6$ EM towers and $5\times10^4$ hadronic towers, and 
suppose that we calibrate $2\times2$ blocks of these as fundamental
units.  We take the calorimeter energy resolutions to be
$\delta E/E=0.15/\sqrt E \oplus 0.005$ (EM) and
$\delta E/E=0.4/\sqrt E \oplus 0.01$ (hadronic).  Then we obtain about 
28 electrons in each EM calibration block and about 24K hadrons in
each hadronic block.   In each block, the precision on the $Z$ mass
measurement is given by $\delta M = \sigma_M/\sqrt N$, with
$\sigma_M/M \approx \delta E/E$ at $E=M_Z$.  The resulting mass measurements
for each block yield a statistical error on the $Z$ mass, and hence
a calibration accuracy of 0.38\% (EM) and 0.03\% (hadronic).  Since these
precisions are small compared with the intrinsic energy resolution 
smearing, they should be adequate for measurements of SM and Susy masses.

It is highly desirable to have calibration methods that can be performed during
high energy running, and thus can track changes in calibration.  Bhabha
scattering has been studied as a means of continuous EM calibration.
Here we focus on an alternate method using `energy flow' (not to
be confused with the technique of substituting charged particle momentum
measurements for calorimeter energy deposits).  
We define energy flow to be the relative number of events seen above
some arbitrary energy threshold, over a ring of cells at fixed azimuth.
The azimuthal symmetry for physics processes means that observed variations
are due to differences in the
 relative gain constants ${\alpha_i}$ for the set
of cells in the ring ($E_{\rm true}=\alpha E_{\rm meas}$).  
A simplified estimate of the calibration precision obtainable can
be made assuming that the energy spectrum in a given azimuthal
ring is given by $dN/dE = Ae^{-BE}$.  Then for $N$ events above
an energy threshold $E_{\rm th}$, one can show that $\delta \alpha
= 1/(BE_{\rm th}\sqrt N)$, and the optimum (smallest $\delta \alpha$) choice
of threshold is $E_{\rm th} = 2/B$.

For 10 \ifb of data at 500 GeV and the
inclusive particle spectrum (we actually used the 
flatter 1000 GeV Monte Carlo
spectrum~\cite{barklow}, and hence
conservative for our purposes), and with 200 cells at fixed $\phi$
for each of 20 bins in $\cos\theta$, we find that a 1\% calibration
can be obtained.  Though of poorer precision than the 
special $Z$ pole calibration, this is a very useful check for drifts
in the calibration during high energy running.  The method itself
can be checked with energy flow measurements at the $Z$ pole, where
we estimate a precision of about 0.8\% for each cell.  Cross calibration
of different azimuthal rings would be performed using $Z$ bosons from
the high energy runs.

The absolute energy calibration of the calorimeter at the
0.03\% level is desired to match the expected statistical precision of the
Higgs boson mass measurement.  This can be obtained after relative
calibration of the calorimeter blocks in the special $Z$ pole runs
discussed above, and should yield calibration of both EM and
hadronic calorimeters to the requisite accuracy. 

\section{Run order}

We should note that before the LC run plan is constructed, 
we will have an understanding of the
actual physics scenario from the LHC/Tevatron.
In particular, the  Higgs mass will be known and the presence
or absence of supersymmetry will be established.  The run order given here
assuming the SM2 Susy scenario discussed above 
is not well tuned, but is offered to guide our
thinking.  Clearly the initial stages will also be constrained
by the progress of the accelerator complex.  We assume the
growth of luminosity in the LC to be that given in
Table~\ref{tab:lumi}.

\begin{table}[t]
\caption{\label{tab:lumi}
Profile by year of the luminosity accumulation.  The luminosity 
is given in \ifb ~assuming 500 GeV operation. 
}
\begin{center}
\begin{tabular}{|cccccccc|}
\hline
Year & 1 & 2 & 3 & 4 & 5 & 6 & 7 \\ \hline
$\int {\cal L}dt$ &10&40&100&150&200&250&250\\ \hline
\end{tabular}
\end{center}
\end{table}

\begin{enumerate}

\item
In the first year, if 500 GeV operation is feasible, it would be profitable to
run there to get a first indication of the masses of the easier
Susy particles to observe.  In the 10 \ifb, we expect $\sim$700 $ZH$ events
at 500 GeV.
If 500 GeV running is not available, operation at 350 GeV would
be called for ($\sim$1500 $ZH$ events).  If it is only possible to 
run at low energy, then one would devote  the initial running
to the $Z$ pole.

\item
Go to 500 GeV as soon as possible for $\sim$80 \ifb to obtain about
two times the ultimate errors on sparticle masses.  This should be
sufficient to establish the $\sell, \selr, \smul, \smur$, and $\chz1$
masses well, and give reasonable estimates for $\chz2, \chpm1$,
and $\stau1$ masses.

\item
Scans at 285 GeV ($\smul \smur$ threshold) and at 350 GeV
($t\overline t, \sell \selr$, and $\chp1 \chm1$ thresholds).

\item
Complete the 500 GeV run for the ultimate precision on end point
masses, and for disentangling the multi-$\tau$ final states.

\item
Scan at 270 GeV ($\chz1\chz2$ and $\stau1 \stau1$ thresholds).

\item
Scan at 285 GeV using $e^-e^-$ initial state to give the precise
$\selr$ mass.

\item
Operate at above the nominal top energy at 580 Gev with \epem to
do the $\chpm1 \chmp2$ study and obtain the $\chpm2$ mass. 

\end{enumerate}

Note that no Giga-Z run is called for our SM2 Susy scenario, since
the need to provide the extra $Z$-pole (and more precise $W$ mass)
is not so great in the scenarios where supersymmetry provides the
mechanism for EWSB.

\section{Considerations for a run scenario if it's not Susy}

If the world we inhabit does not include low mass
supersymmetry, we will know this at the start of the LC operation.
In this world, there are two avenues for exploring the character
of EWSB:  observe new states or deviations in cross-sections
from SM predictions at high energy; and improve our knowledge
of the precision electroweak observables at the $Z$ pole
and $WW$ threshold.  The run plan in this case should reflect
these two paths, and we outline such a plan below. 
Approximate equivalent luminosity acquisitions (the accumulations
that would have occured in the same time at  500 GeV) are given
in parentheses.  We have not specified the order in which the runs
might be taken, but clearly those involving very careful understanding
of the beam energies (for the $WW$ threshold run), positron polarization
(for Giga-$Z$), or $\gamma \gamma$ collisions should be left for
later in the program when the machine is well-understood.

\begin{enumerate}

\item
(500 \ifb)  Look for new high mass phenomena

\item
(150 \ifb) Run at the highest accessible energy by trading off
luminosity for energy.  Seek new signals at higher energies
and new physics energy dependences.

\item
(100 \ifb)
Obtain the full Giga-$Z$ sample, with polarized $e^+$.  About
20 \ifb are needed at the $Z$ (thus ${\cal L}_{\rm equiv} \approx 100$ \ifb).

\item
(150 \ifb)
Run with $\gamma \gamma$ collisions at the maximum energy
available to give an alternate view to \epem collisions
for the trilinear gauge couplings, heavy $Z^\prime$, or complementary
sensitivity to large extra dimensions.

\item
(80 \ifb)
Scan the $WW$ threshold for a precision $W$ boson mass
measurement.

\item
(20 \ifb)
Scan the $t\overline t$ threshold for precision top quark
mass and width information.

\end{enumerate}

\section{Conclusions}

In the delightful case that Nature provides us with low mass supersymmetry,
the LC should offer the possibility for extensive and precise
measurements to point the way for understanding EWSB
and the character of physics beyond the SM.  The details of the run
plan will depend critically upon the exact 
model that we encounter.  Considerable ingenuity may be required to unscramble
the effects of many accessible supersymmetric particles, but in the 
$\tau$-rich case examined here, it seems possible to do a good job with this.
Even for the low mass Susy situation studied here, much running at 500 GeV is
needed to establish a good picture of the various sparticle masses.  And it
remains clear that further studies at energies above  500 GeV will be needed
in this, or other conceivable physics scenarios.

\section*{Acknowledgments}
I am grateful to G. Bernardi, G. Blair, J. Butler, 
R. Cahn, J.K. Mizukoshi, 
U. Nauenberg and G.W. Wilson for important contributions
to the studies reported here.
This work was supported by the US National Science Foundation, grant
0096707.


\section*{References}
\begin{thebibliography}{99}

\bibitem{snowmass}M. Battaglia {\it et al.}, Proceedings of Snowmass 2001, 
eConf C010630, SLAC-R-599, ed. Norman Graf, paper E3006
(hep-ph/0201177).

\bibitem{allanach} B.C. Allanach, hep-ph/0202233.

\bibitem{martyn} H.U. Martyn and G. Blair,
hep-ph/9910416.

\bibitem{nauenberg} U. Nauenberg {\it et al.}, 
Proceedings of Snowmass 2001, 
eConf C010630, SLAC-R-599, ed. Norman Graf, paper E3056.

\bibitem{mizukoshi} J.K. Mizukoshi, H. Baer, A.S. Belyaev, X. Tata,
hep-ph/0107216.

\bibitem{blair} G. Blair, 
Proceedings of Snowmass 2001, 
eConf C010630, SLAC-R-599, ed. Norman Graf, paper E3019.

\bibitem{feng} J. Feng and M. Peskin, 
hep-ph/0105100.

\bibitem{tesla} ``{\it TESLA Technical Design Report}'', DESY 2001-011,
(March 2001), 
http://tesla.desy.de/new\_pages/TDR\_CD/start.html. 

\bibitem{orange} ``{\it Linear Collider Physics Resource
Book for Snowmass 2001 }'', American Linear Collider Working Group,
SLAC-R-570,  
http://www.slac.stanford.edu/grp/th/LCBook/.

\bibitem{barklow} T. Barklow, private communication.

\end{thebibliography}
\end{document}